\documentclass[12pt]{elsart}

\usepackage{epsf}

\newcommand{\PSfigure}[2]{%
    \centerline{\epsfxsize=#2\textwidth\epsffile{#1}}
}
\newcommand{\ba}{\begin{eqnarray}}
\newcommand{\ea}{\end{eqnarray}}
\newcommand{\be}{\begin{equation}}
\newcommand{\ee}{\end{equation}}

\begin{document}

\begin{frontmatter}

\title{Left-Handed Surface Waves in a Photonic Structure }

\author[IIT]{G.~Shvets},

\address[IIT]{
    Center for Accelerator and Particle Physics, Illinois Institute of
    Technology, Chicago IL 60302, USA }

\thanks{%
    This work supported by the US DOE Division of
    High Energy and Nuclear Physics and the Presidential
    Early Career Award for Scientists and Engineers
}


\begin{abstract}
It is demonstrated that an isotropic left-handed medium can be
constructed as a photonic structure consisting of two dielectric
materials, one with positive and another with negative dielectric
permittivities $\epsilon$. Electromagnetic waves supported by this
structure are the surface waves localized at the dielectric
interfaces. These surface waves can be either surface phonons or
surface plasmons. Two examples of negative $\epsilon$ materials
are used: silicon carbide and free-electron gas.
\end{abstract}

\end{frontmatter}
\section{Introduction}
Left-handed materials (LHMs)~\cite{veselago} with negative
magnetic permeability $\mu$ and dielectric permittivity $\epsilon$
have recently attracted a great deal of attention because of their
promise for developing novel lenses and low reflectance
surfaces~\cite{pendry_lens,smith_kroll}. LHMs have a negative
refractive index, which implies that the phase and group
velocities of the propagating electromagnetic wave oppose each
other. This property of LHMs is responsible for their anomalous
physical behavior: reversed sign of the Doppler effect and
counter-intuitive refraction at the interface with the ''normal''
(right-handed) media~\cite{veselago}.

Since materials with negative refraction index do not naturally
occur, they have to be artificially constructed. An LHM in the
microwave frequency band was recently constructed as an array
consisting of metal rods and split-ring resonators
~\cite{smith_prl00}. Using the same strategy for designing LHMs at
much shorter wavelengths (e.~g., in the infrared part of the
spectrum) may prove challenging. Here I describe a new idea of
making a LHM {\it photonic} structure by utilizing surface waves
at the interfaces of dielectrics with opposing signs of
$\epsilon$. Left-handedness of such structures is due to the
existence of the surface waves at the vacuum/dielectric
interfaces. Practical implementation of such photonic structures
at a micron scale is encouraged by the availability of low-loss
dielectrics and semiconductors with negative $\epsilon$, including
many polar crystals such as SiC, LiTaO$_3$, LiF, and ZnSe. The
frequency-dependent dielectric permittivity of these crystals,
given by the approximate formula~\cite{kittel} $ \epsilon(\omega)
= \epsilon_{\infty} (\omega^2 - \omega_L^2)/(\omega^2 -
\omega_T^2),$ is negative for $\omega_T < \omega < \omega_L$. Note
that $\epsilon(\omega)$ reduces to that of the free electron gas
for $\epsilon_{\infty} = 1$ and $\omega_T = 0$.

In Section~\ref{sec:single_waveguide} I review the properties of
surface waves in a single waveguide (SW) with $\epsilon_c < 0$
cladding shown in Fig.~\ref{fig:figure_structures}(a), derive the
effective dielectric permittivity $\epsilon_{\rm eff}$ and
magnetic permeability $\mu_{\rm eff}$, and demonstrate that the
phase and group velocities of the surface waves can oppose each
other. Waveguide with silicon carbide (SiC) cladding is used as an
example of a medium which supports left-handed surface phonons.

In Section~\ref{sec:triang_lattice} the triangular lattice
photonic waveguide (TLPW) shown in
Fig.~\ref{fig:figure_structures}(b) is found to be a perfectly
isotropic LHM for small wavenumbers $k_{x,y} \ll \pi/d$, where $d$
is the lattice periodicity. The triangular cladding regions shown
in Fig.~\ref{fig:figure_structures}(b) are assumed to have
$\epsilon = 1 - \omega_p^2/\omega^2$ with the frequency dependence
of a free electron gas. In this example, vacuum/cladding
interfaces support left-handed surface plasmons.
Recently~\cite{valanju}, it has been suggested that the group
velocity in LHMs is not aligned with the phase velocity, making
the perfect lens~\cite{pendry_lens} impossible. I demonstrate that
this is not the case even for an artificially constructed LHM such
as the TLPW.

\section{Single waveguide with negative $\epsilon$ cladding}
\label{sec:single_waveguide}

To illustrate how a negative $\mu$ can be mimicked in a single
waveguide, consider electromagnetic wave propagation in the
horizontal ($x$) direction assuming a piecewise constant
dielectric constant: $\epsilon = 1$ in the vacuum channel (for $-b
< y < b$) and $\epsilon = \epsilon_c < 0$ inside the cladding (for
$|y| > b$). The goal is to explain how the energy can flow in the
negative $x-$direction (and so $\partial \omega/\partial k_x < 0$)
whereas the phase velocity $\omega/k_x > 0$. Consider a confined
transverse magnetic (TM) wave with non-vanishing components $(E_x,
E_y, H_z) \propto \exp{i(k_x x - \omega t)}$ such that $E_y = H_z
= \partial_y H_z = 0$ at midplane $y = 0$. Since we are interested
in the wave propagation along $x$, introduce integrated over the
transverse direction $y$ quantities
\[
    \tilde{E}_{x,y} = \int_0^{\infty} dy \ E_{x,y}, \ \ \ \
    \tilde{H}_{z} = \int_0^{\infty} dy \ H_{z}.
\]
From Faraday's and Ampere's laws, assuming that $E_x(x \rightarrow
\infty) = 0$ and integrating by parts, obtain, correspondingly,
$\partial_x \tilde{E}_y = i \omega \mu_{\rm eff} \tilde{H}_z/c$
and $\partial_x \tilde{H}_z = i \omega \epsilon_{\rm eff}
\tilde{E}_y/c$, where the effective dielectric permittivity and
magnetic permeability of the waveguide are defined as follows:
\begin{equation}
    \epsilon_{\rm eff} = \frac{1}{\tilde{E}_y} \int_0^{\infty} dy
    \epsilon E_y, \ \ \ \mu_{\rm eff} = 1 + \frac{i c}{\omega}
    \frac{E_x(y=0)}{\tilde{H}_z}.
    \label{eq:epsilon_eff}
\end{equation}

The definition of the weight-averaged $\epsilon$ is intuitive, and
$\mu_{\rm eff}$ is defined so as to eliminate the longitudinal
component of the electric field $E_x$ which does not contribute to
the power flow along the waveguide. For $k_x > 0$ note that $E_y$
and $B_z$ are, on average, out of phase if $\mu_{\rm eff} < 0$,
making the wave left-handed. The resulting dispersion relation is
$k_x^2 c^2/\omega^2 = \mu_{\rm eff} \epsilon_{\rm eff}$.
Therefore, $\epsilon_{\rm eff}$ and $\mu_{\rm eff}$ must be of the
same sign for a propagating wave. The propagating mode is
left-handed if $\mu_{\rm eff} < 0, \epsilon_{\rm eff} < 0$,
necessitating that the dielectric constant of the cladding be
negative.

\begin{figure}
    \PSfigure{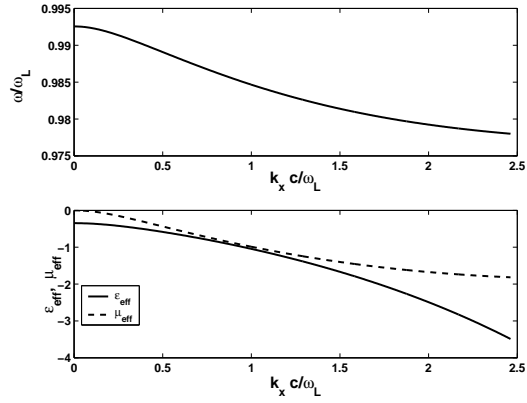}{0.5}
    \caption{(a) Dispersion
    relation, and (b) effective dielectric permittivity $\epsilon_{\rm
    eff}$ and magnetic permeability $\mu_{\rm eff}$. Dielectric
    permittivity of cladding $\epsilon_c = 1 - \omega_p^2/\omega^2$,
    gap width $2b = c/\omega_p$.}
    \label{fig:1dchannel}
\end{figure}

$\epsilon_{\rm eff}$ and $\mu_{\rm eff}$ are calculated by solving
the eigenvalue equation for $H_z$:
\begin{equation}
    \frac{\partial}{\partial x} \left( \frac{1}{\epsilon}
    \frac{\partial H_z}{\partial x} \right) + \frac{\partial}{\partial y}
    \left( \frac{1}{\epsilon} \frac{\partial H_z}{\partial y} \right) =
    - \frac{\omega^2}{c^2}  H_z,
    \label{eq:by1}
\end{equation}
where $\partial_x = i k_x$. As an example, consider silicon
carbide (SiC) cladding with $\epsilon_{\infty} = 6.7$, $\omega_L =
182.7 \times 10^{12} {\rm s}^{-1}$, and $\omega_T = 149.5 \times
10^{12} {\rm s}^{-1}$~\cite{spitzer_pr59}. SiC has a stopband
(region of negative $\epsilon$) in the far infrared, $10.3  <
\lambda < 12.6 \mu$m. The dispersion relation and the
corresponding $\epsilon_{\rm eff}$ and $\mu_{\rm eff}$ are plotted
in Fig.~\ref{fig:1dchannel}(a,b) for a SW with the vacuum gap
width $2b = c/\omega_p$. The propagating surface mode in a SW
exhibits left-handedness: it's group velocity $v_g = \partial
\omega/\partial k < 0$ is negative, and so are $\mu_{\rm eff} < 0$
and $\epsilon_{\rm eff} < 0$. The cutoff at $\omega = 0.993
\omega_L$ is caused by the vanishing of the $\mu_{\rm eff}$.

Why is $v_g < 0$ despite $v_{\rm ph} = \omega/k > 0$? The total
Poynting flux $P_x = c E_y H_z/4\pi$ along the dielectric
waveguide is the sum of the fluxes inside the cladding and in the
vacuum gap. In the gap, $E_y$ and $H_z$ are in phase, so $P_x >
0$. Inside the cladding $E_y$ reverses its sign across the
vacuum/cladding interface because $\epsilon_c < 0$ (continuity of
$D_y = \epsilon E_y$). Because $H_z$ is continuous across the
interface, $P_x < 0$ in the cladding. For a narrow gap, the
integrated Poynting flux is negative. Note that not any surface
wave is left-handed. Achieving left-handedness requires that (a)
the frequency of the mode lie within the stopband of the cladding,
(b) there are two interfaces, and (c) the vacuum gap between the
interfaces is small. In the case of SiC cladding, the left-handed
waves are surface phonons.

\section{Triangular Lattice Photonic Waveguide}
\label{sec:triang_lattice}

\begin{figure}
    \PSfigure{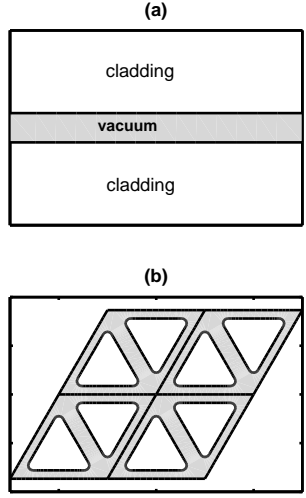}{0.5}
    \caption{(a)
    Single waveguide (SW) consists of a vacuum gap
    surrounded by cladding with $\epsilon_c < 0$. (b)
    Triangular lattice photonic waveguide (TLPW): dielectric cladding regions (white
    triangles) separated by vacuum channels (shaded).}
    \label{fig:figure_structures}
\end{figure}

The single waveguide example was used to explain the emergence of
left-handedness of the surface waves, and to derive the necessary
conditions (a-c) for their existence. The objective of this paper
is to construct an isotropic photonic medium capable of
transmitting left-handed waves in all directions. This is
accomplished by constructing a Triangular Lattice Photonic
Waveguide (TLPW) which consists of the triangular pieces of
dielectric cladding arranged in a triangular lattice, as shown in
Fig.~\ref{fig:figure_structures}(b). We assume that the dielectric
permittivity of the cladding is the same as that of a free
electron gas: $\epsilon_c = 1 - \omega_p^2/\omega^2$.

\begin{figure}
    \PSfigure{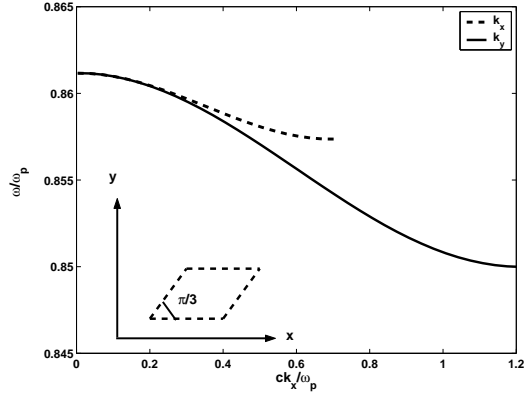}{0.5}
    \caption{Dispersion relation $\omega$ v.~s. $|\vec{k}|$ for a
    triangular arrangement of dielectrics as in
    Fig.~\ref{fig:figure_structures}(d). Equilateral parallelogram
    with $d = 3 c/\omega_p$ and $\beta = \pi/3$ opening angle --
    elementary cell of the photonic structure. Solid line: $\vec{k} =
    k \vec{e}_y$, $0 < kd < 2 \pi/\sqrt{3}$; dashed line: $\vec{k} = k
    \vec{e}_x$, $0 < kd < 2 \pi/3$. Channel widths $2b = 0.6
    c/\omega_p$, dielectric edges smoothed with radius $r_b = b$.}
    \label{fig:triang_band}
\end{figure}

Every propagating wave in the TLPW is characterized by its
wavenumber $\vec{k}_{\perp} = k_x \vec{e}_x + k_y \vec{e}_y$, and
can be represented as $H_z = \tilde{H} \exp{i \vec{k}_{\perp}
\cdot \vec{x}}$. Here $\tilde{H}$ is periodic on the boundaries of
the elementary cell of the triangular lattice shown as an
equilateral parallelogram in Fig.~\ref{fig:triang_band}. For each
value of $\vec{k}_{\perp}$ Eq.~(\ref{eq:by1}) is solved for its
eigenvalues using a commercial finite elements code
FEMLAB~\cite{ref_femlab}, yielding the dispersion relation
$\omega$ v.~s.~$\vec{k}_{\perp}$. The local $\vec{P}(x,y) = c
\vec{E} \times \vec{H}/4\pi$ and cell-averaged $\vec{P}_{\rm av} =
\langle \vec{P}(x,y) \rangle $ Poynting fluxes are computed for
each solution.

Simulations results indicate that TLPW is perfectly isotropic for
small $\vec{k}$ well inside the Brillouin zone. The elementary
cell of the TLPW and the dispersion relation $\omega$
v.~s.~$\vec{k}$ for two orthogonal directions of $\vec{k} =
\vec{e}_{x,y} k$ are shown in Fig.~\ref{fig:triang_band}. The two
dispersion curves, which are identical for $|\vec{k}| \ll \pi/d$
(establishing the isotropy), are drawn to the respective edges of
the Brillouin zone: $0 < k_x d < 2\pi/3$ and $0 < k_y d <
2\pi/\sqrt{3}$~\cite{smirnova_jap}.

\begin{figure}
    \PSfigure{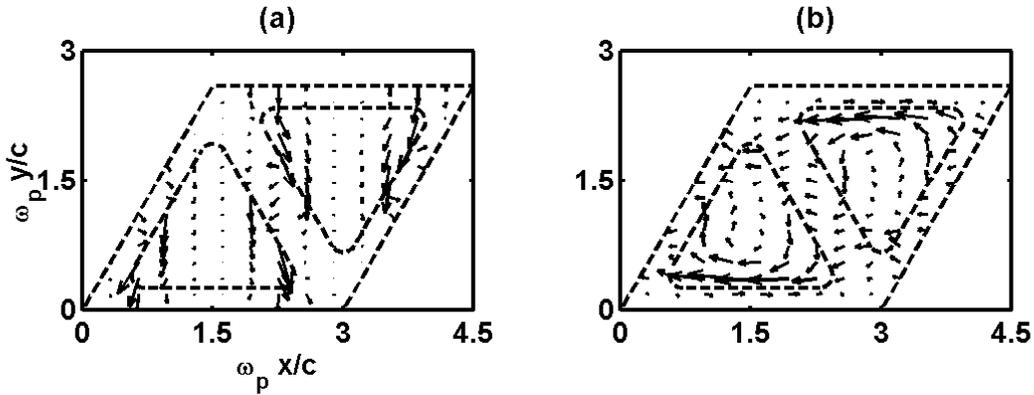}{1.0}
    \caption{Power flow in a Triangular Lattice Photonic Waveguide.
    Structure parameters same as in Fig.~\ref{fig:triang_band}.
    Wavenumber $k = d^{-1} \pi/6$, and (a) $\vec{k} = k \vec{e}_y$ (no
    vacuum/cladding interfaces parallel to $\vec{k}$); (b) $\vec{k} =
    k \vec{e}_x$, (interface along $\vec{k}$) }
    \label{fig:flow_kxky}
\end{figure}

These two directions are chosen because they exhibit the maximum
possible anisotropy. Indeed, for $\vec{k} = k \vec{e}_x$ there
exists a vacuum/cladding interface along which the electromagnetic
energy can flow in $x-$direction, Fig.~\ref{fig:flow_kxky}(b). No
such interface exists for $\vec{k} = k \vec{e}_y$,
Fig.~\ref{fig:flow_kxky}(a). Thus, the local Poynting flux
patterns are very different for these two directions, as can be
seen by comparing Figs.~\ref{fig:flow_kxky}(a,b). For the
parameters of Fig.~(\ref{fig:flow_kxky}), $k = d^{-1} \pi/6$, the
index of refraction is $n = -ck/\omega = -0.2$. Index of
refraction can be readily tuned in both directions by adjusting
the parameters of the TLPW: periodicity $d$ and channel width
$2b$.

Despite the differences in the flow patterns, the cell-integrated
fluxes $\vec{P}_{\rm av}$ are identical for all directions of $\vec{k}$,
and so are the frequencies: $\omega = 0.86 \omega_p$. Numerical results
unambiguously confirm that, for small $|\vec{k}|$, phase and group
velocities exactly oppose each other. At least for this particular
LHM, the claim of Valanju~{\it et.~al.~} \cite{valanju} that
$\alpha = \angle (\vec{v}_{\rm ph}, \vec{v}_{\rm g}) \neq \pi$ is
not confirmed.  Anisotropy for large $|\vec{k}|$ is merely the
consequence of the periodicity of the photonic structure.

In conclusion, it was demonstrated that surface waves (phonons or
plasmons) at the interface of dielectrics with opposing signs of
the dielectric permittivity $\epsilon$ can be left-handed. An
isotropic composite medium which has a negative refraction index
with respect to such waves can be constructed as a photonic
structure.

\end{document}